# Construction of an NP Problem with an Exponential Lower Bound

Roman V. Yampolskiy
Computer Engineering and Computer Science
University of Louisville
roman.yampolskiy@louisville.edu

**Abstract**
In this paper we present a Hashed-Path Traveling Salesperson Problem (HPTSP), a new type of problem which has the interesting property of having no polynomial time solutions. Next we show that HPTSP is in the class NP by demonstrating that local information about sub-routes is insufficient to compute the complete value of each route. As a consequence, via Ladner's theorem, we show that the class NPI is non-empty.

**Introduction**
In a recent survey of 100 complexity theorists, a great majority stated that they believed that P ≠ NP, but a surprising 9 scientists thought that it will be shown that P = NP [9]. The opinion of the majority is easy to justify, a proof of P ≠ NP would preserve the status quo, be consistent with our intuitive notions of difficulty, and explain why over 50 years of search for the polynomial time algorithm to solve NP-Complete problems has not produced any results [7, 6, 5, 8, 12, 13, 15]. On the other hand, if P = NP, the world would be a very different place [10]. Difficult problems like factoring [17] would be easy, privacy (at least in the digital sense) would not exist, NP would have to equal CoNP, and many other unlikely relationships between complexity classes will also follow.

So why do some respected scientists believe that P = NP? For some, no doubt, it is just intellectual honesty; since we have no proof in either direction we should leave both possibilities on the table. But for many in the "P could equal NP" camp the properties of the NP problems are exactly what make them believers. For one, the quick verification of the potential solutions may hold the key to quick determination of solutions. Also, ever better polynomial time approximation algorithms are steadily developed, perhaps one day we will get so close we are actually there. But the most important reason for hope, in our opinion, is the ability to evaluate and compare partial information about the solutions to NP-Complete problems and by doing so, prune away some of the potential solutions in the exponential search space.

Let's look at a particular example - the infamous Traveling Salesman Problem (TSP) [18]. Any polynomial time algorithm for solving the TSP would have to prune away a proportionate majority of the exponential number of potential paths without ever examining them. The only way to cancel out a potential path in TSP without explicitly examining it, is to analyze its comprising components and based on their properties conclude that the complete path will not be competitive with respect to already examined full paths or other components. The information about distances between any two cities is provided to us. For any potential TSP edge or sub-path, we can obtain its numeric value and compare it to other edges or paths/sub-paths. We can freely examine the instance of TSP provided to us and determine for example that a particular edge is so much longer than any other that including it in the path will guarantee that that path is not the shortest. As the result, we can prune away (never explicitly examine) all paths which contain this undesirable edge. And so, potentially, by only looking at a polynomial number of local segments we can make drastic reductions in the size of the total search space.

It is important to note that paths with similar properties (length) are often located adjacently to each other in lexicographic and other types of ordering in which paths could be considered. This makes it possible to address whole sub-regions of the search space with just a sequential numbering of the first and last paths in a sub-group of similar routes, bypassing the need to explicitly list each of exponentially many paths we want to prune. Consequently, the ability to evaluate and compare small segments of the solution in some very intelligent way may give us a polynomial time algorithm for solving TSP, or so the logic goes.

The situation we described above is not unique to TSP and is trivially a property of all NP-Complete problems. To remove all doubt that a particular NP problem does not have a polynomial solution we need to remove the possibility of pruning away potential solutions without examining them individually. This can be accomplished if the ability to obtain information by examining local substructures of the potential solutions is not present in a problem. In the next sections we review necessary tools, and present an artificial problem which has this desirable property. We begin by reviewing the concept of hashing.

**The Hash Functions**

A hash function takes a string and returns a fixed-size string, the hash value, such that any change to the data will change the hash value. Figure 1 shows performance of a hash function SHA-1 the most popular hash function currently available. SHA stands for Secure Hash Algorithm and was designed by the National Security Agency and published by the NIST as a U.S. Federal Information Processing Standard. SHA-1 produces a 160-bit hash value. Figure 2 illustrates an iteration of SHA-1 compression function.

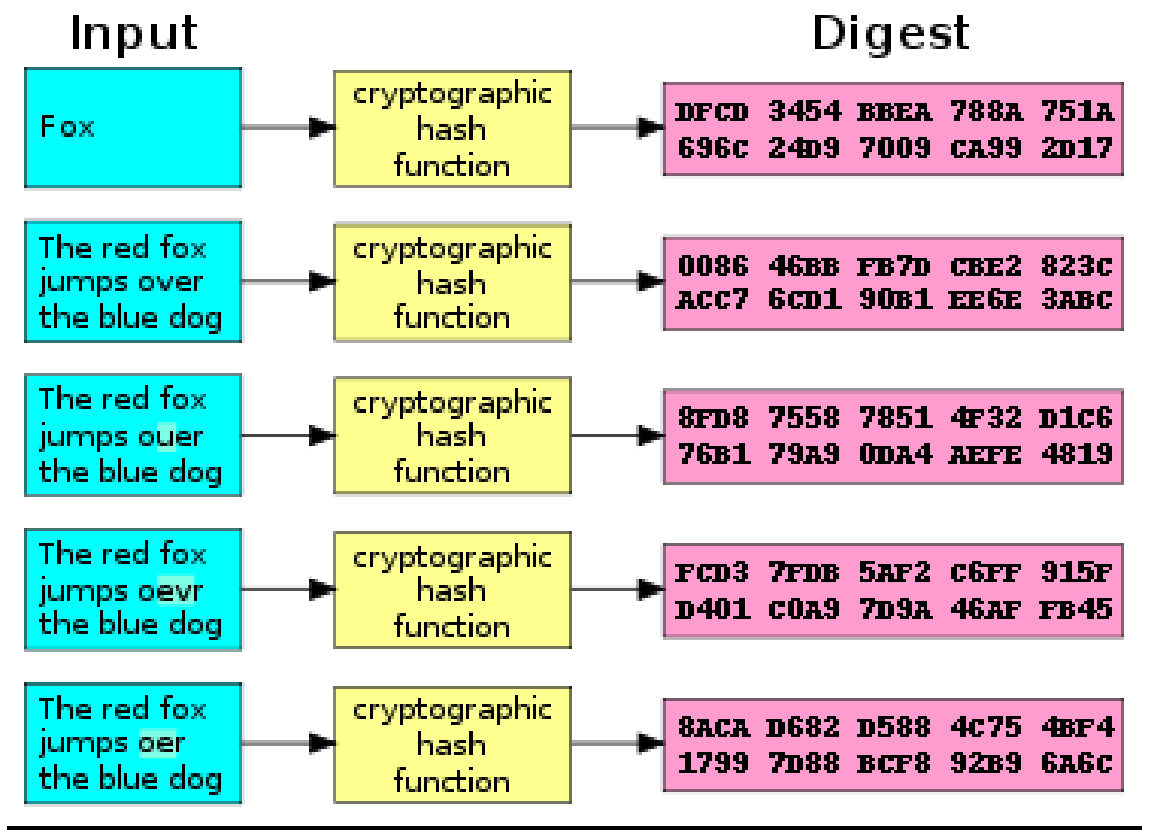

Figure 1: Sample output from SHA-1 varied with respect to the input [2].

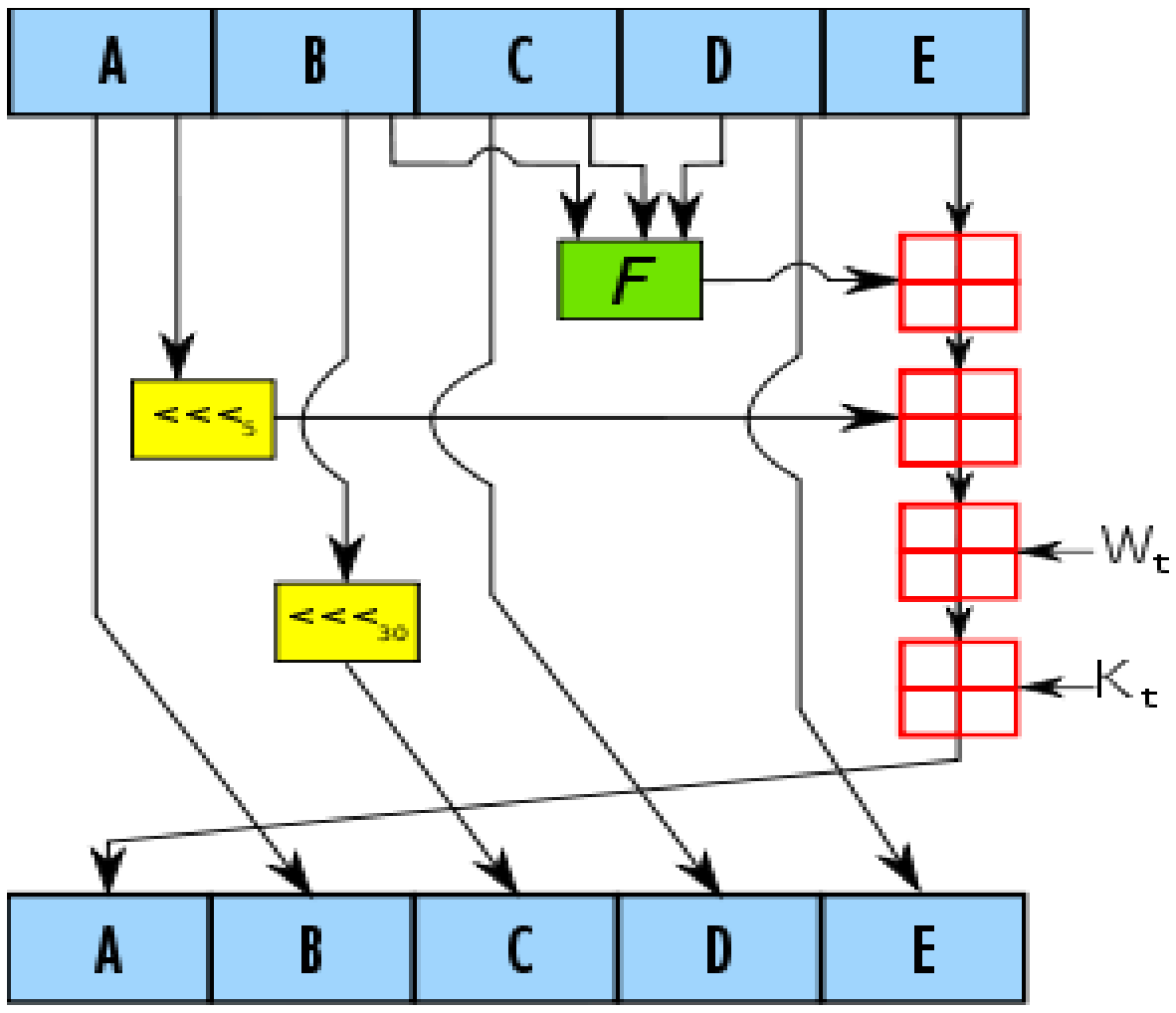

Figure 2: Iteration of the SHA-1 compression function: A, B, C, D and E are 32-bit words of the state; *F* is a nonlinear function that varies; $\lll_n$ denotes a left bit rotation by *n* places; *n* varies for each operation; $W_t$ is the expanded message word of round t; $K_t$ is the round constant of round t; $\boxplus$denotes addition modulo $2^{32}$ [4].

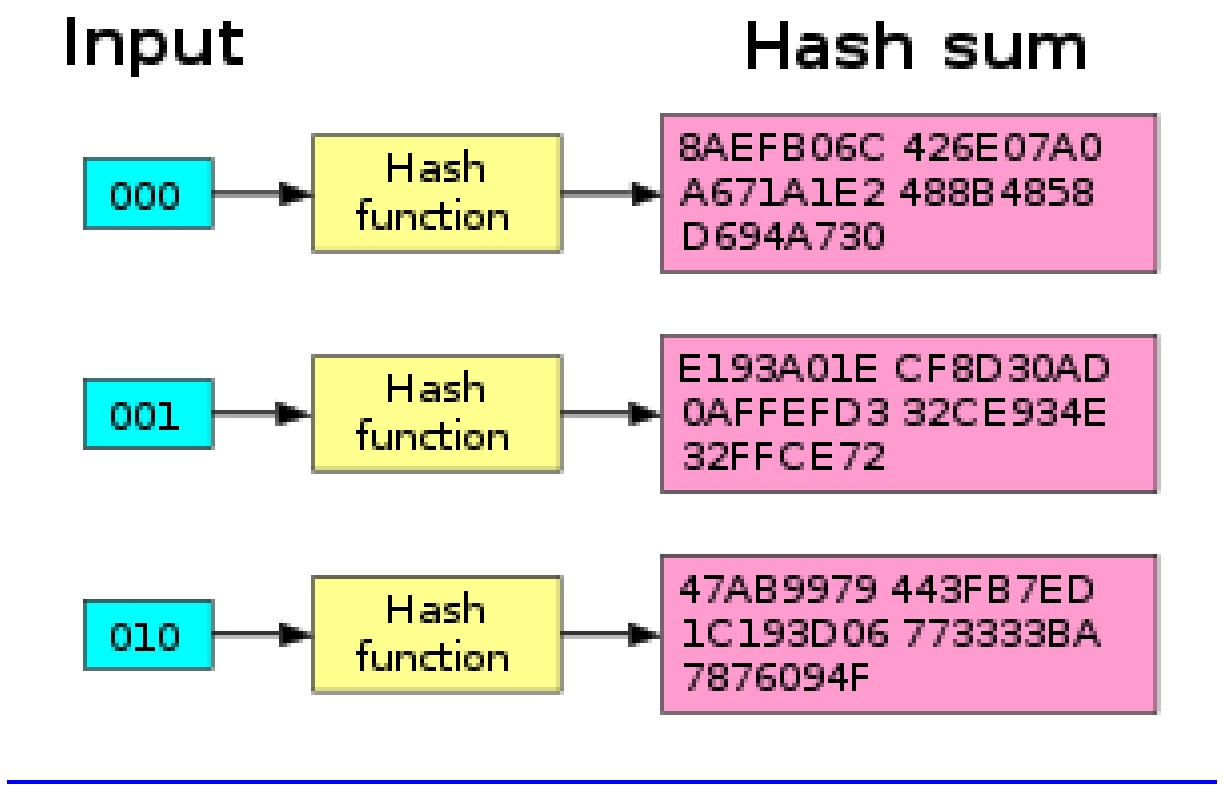

Figure 3: Avalanche effect in action [1].

One of the most desirable properties of SHA-1 is a strong avalanche effect. If a single bit of input is changed the hash value becomes completely different, resulting in good randomization of the output string, as shown in Figure 3. The Strict Avalanche Criterion (SAC) is a generalization of the avalanche effect, whenever a single input bit is flipped, each of the output bits changes with a probability of 50%. The SAC was introduced by Webster and Tavares [14]. A hash function must be able to process a string of any length into a fixed-size output. This can be achieved by breaking the input up into a sequence of blocks, and operating on them in order. Figure 4 illustrated the Merkle–Damgård construction used for such purposes.

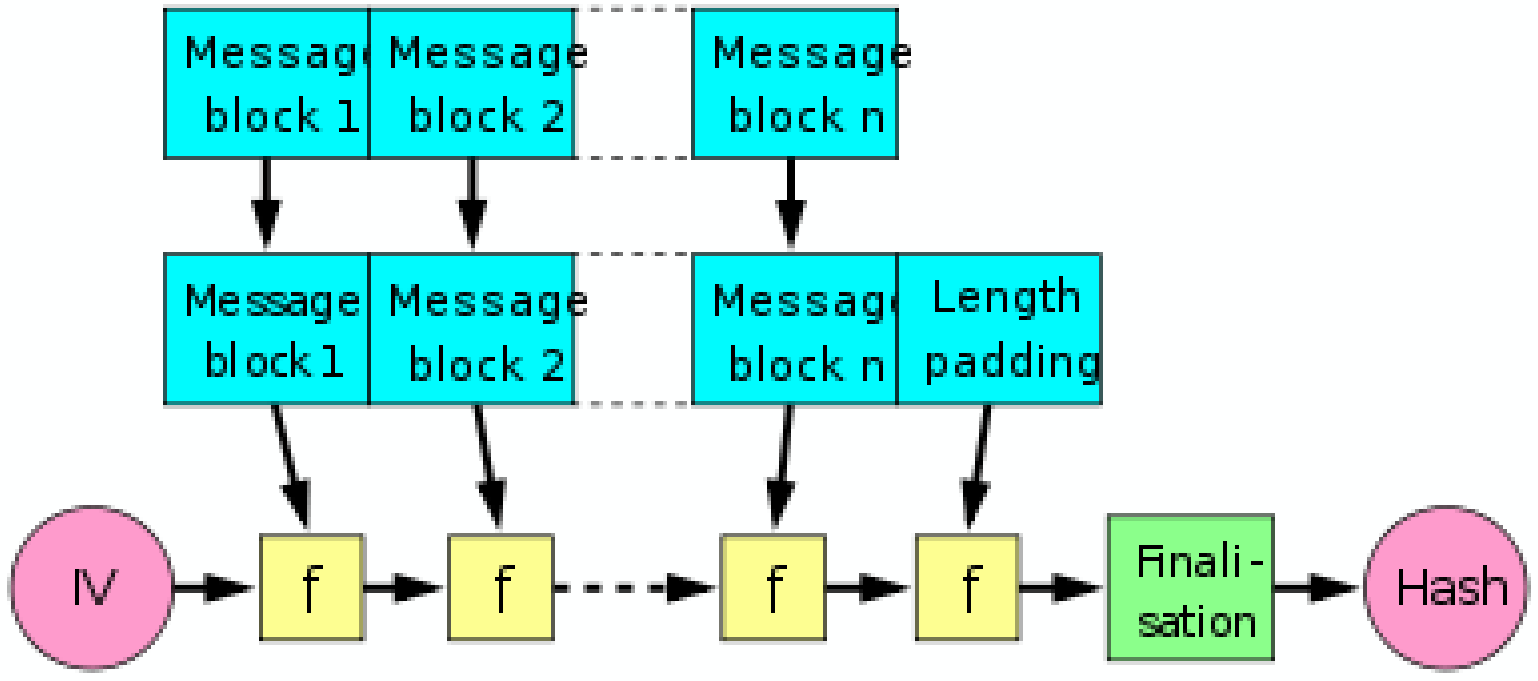

Figure 4: The Merkle–Damgård hash construction [3].

**Hashed Path TSP (HPTSP)**

We propose a novel artificial problem with the desirable property of providing no local information about its potential solutions, and consequently preventing pruning of the search space. The proposed problem is a variant of the classical TSP problem presented next. *Traveling Salesman Problem:* G = (V,E) is a complete graph. All edges have costs. Problem: Find a min-cost tour (Hamiltonian Cycle). TSP could be formalized as follows:

*TSP = { <G, c, k> : G = (V, E) is a complete graph,*
*c is a cost function V x V → Z,*
*k ∈ Z, and G has a tour with cost ≤ k }*

The proposed Hashed-Path TSP (HPTSP) takes an existing TSP problem instance and instead of asking to find a min-cost route asks: What is the route (with weights included) with a lowest lexicographical order after the strings representing the paths have been hashed. The following example should explain both construction and evaluation of an HPSTP instance. Let's say we start with the TSP instance depicted in Figure 5.

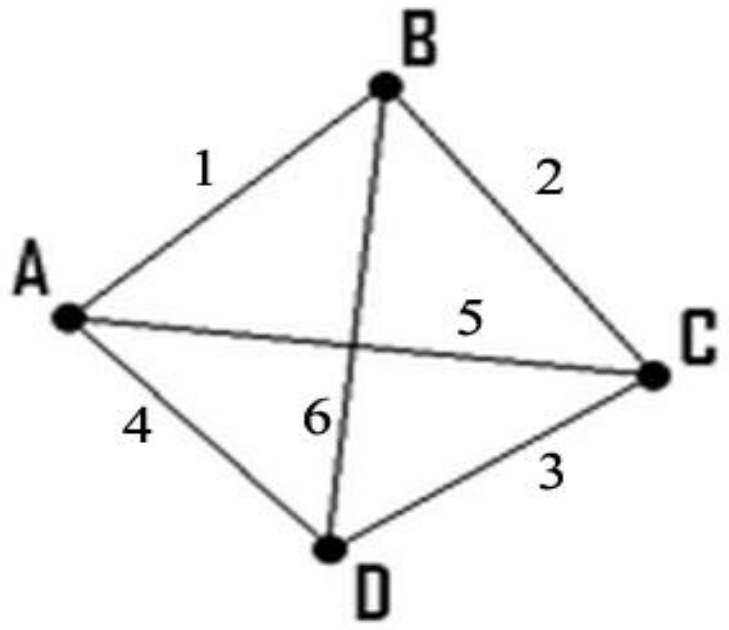

Figure 5. 4 city TSP example.

The corresponding instance of HPTSP would be exactly the same, but the question we ask is completely different. We would like to find a path with the lowest lexicographical order after the strings representing the paths have been hashed. Just like with TSP, where a specific value of k could be given, a specific lexicographic constant could be provided to HPTSP. We will now show an exponential time algorithm for solving this trivial instance of HPTSP as a way to further explain the proposed problem.

The complete list of paths for this instance is shown in Table 1 under "Routes": For our purposes the full route is the list of nodes with weights in between them, as such: A1B2C3D4. The complete set of full routes is given in the middle column of Table 1.

| Routes | Full Routes | Hashed Full Routes (SHA-1) |
|---|---|---|
| ABCD | A1B2C3D4 | 897ca6fcdeed5883fd7bd85eae55406ac81d9d74 |
| ABDC | A1B6D3C5 | 2bc9a32bd7d7898bbb7ac99fdcc23d3891133c08 |
| ADBC | A4D6B2C5 | a89b60d452e3c315f456f2095516695cb70abd61 |
| DABC | D4A1B2C3 | d315c57519eda431bdc382a5b80e72e22b6e1ec7 |
| ACBD | A5C2B6D4 | 28a6a5675fdb532f01adfb98283463808f99008d |
| ACDB | A5C3D6B1 | 1d3eadcc4495af0d49be4a05181b46375743e4f1 |
| ADCB | A4D3C2B1 | 981cfa7c4a720a33fb6935b7d40863b16723f26a |
| DACB | D4A5C2B6 | aaa25da8ee5d45643185901587ff2916a807aa48 |
| CABD | C5A1B6D3 | 4a0a19a780c292d3fb11be5e79f113f7467a9b18 |
| CADB | C5A4D6B2 | 4873e0be21d8d061e6c58ad0cc2c3190108884bd |
| CDAB | C3D4A1B2 | 73fa14f1c14e67dffae49aa19dde0a2754d67c3a |
| DCAB | D3C5A1B6 | **0274a90142fff8495ee8fc6309bbea1abe6fe9db** |
| BACD | B1A5C3D6 | 4ee28c970941442744512cedeff440a3292f6c53 |
| BADC | B1A4D3C2 | 56ed850a77ee2042c5d0994457e6bd24ae1c9920 |
| BDAC | B6D4A5C2 | 38d6c1b303a5adccc0b94d040df2c500affd8c1d |
| DBAC | D6B1A5C3 | 67781a7d43eb3c697dae423d7ac65b276b8f1cfd |
| BCAD | B2C5A4D6 | a0e5014b5714b581faf87e0e19060d18bec3e97e |
| BCDA | B2C3D4A1 | b2c86b8344548157bbfa294368d6b3a2cb01ec41 |
| BDCA | B6D3C5A1 | 9791f10967c97732cdec796911c6a96986f12985 |
| DBCA | D6B2C5A4 | 9d6b2df207612a83f6809ae9642894eb706ecc53 |
| CBAD | C2B1A4D3 | 0f0b6126cf3663d128ee2d0d7298922c36c13adf |
| CBDA | C2B6D4A5 | a7a4a0c5071801d987e2e3178de47275e035be9a |
| CDBA | C3D6B1A5 | 2d46e345d8cea638bd3dbb4ee1b5c85b1473a340 |
| DCBA | D3C2B1A4 | ba2a8b3e5f4054015c82bec30f03392765e24720 |

Table 1: Routes, full routes and hashed full route values for a given 4 city HPTSP.

Next we select a hash function which has the following properties:

- Strict Avalanche Criterion - the value of each output bit depends on all input bits. Whenever a single input bit is flipped, each of the output bits changes with a probability of 50%.
- Accepts messages of any size and produces a fixed length hash value.

It is important to note that most modern hash algorithms have the desired properties and function in polynomial time. In this particular example we used the SHA-1 algorithm. Next we hash each of the full paths calculated above and get the strings presented in Table 1 under "Hashed full routes" (in hexadecimal). Next we go through the list of paths keeping the one with the lowest lexicographical order found so far. Once the list is exhausted, we have found our solution. In this case the answer to the given instance of HPTSP with respect to the lowest lexicographical order is: "0274a90142fff8495ee8fc6309bbea1abe6fe9db" which is generated from the path DCAB.

We propose the following formalization of HPTSP:

HPTSP = {<G, h, m> : G = (V, E) is a complete graph. h is a hash function. m is a lexicographical constant. There is a route $h(z_i)$ (for i from 1 to V!) with lexicographical value $\leq$ m}

**HPTSP ∈ NP**

*Certificate:* sequence of |V| vertices with edge costs inserted.

*Algorithm:*

Check that vertices don't occur more than once, all vertices are included and the edge costs are correct and in proper locations. Using the hash function, obtain the value of the certificate. Check that it has lexicographical value less than m.

*Polynomial time:*

O(V) to verify format of the certificate.
O(V) to get edge costs.
O(V) to perform hashing (Hashing function itself is of constant size).
O(V) to confirm lexicographical ordering.

**HPTSP is not in P**

Any conceivable polynomial time algorithm for solving HPTSP needs to discard a large proportion of routes without examining them. This needs to happen under the constraint of examining at most a polynomial number of local sectors, paths or their interactions.

The novelty of our approach is in separating information about the parts of the routes from the values of the whole solutions. In the TSP case, each edge contains information about its length, a value which could be compared to other edges or even full paths to make decisions about pruning. In the case of the HPTSP edge information leaks no information about the value of the total solution. The same edge will have a completely different value after encoding, depending on the location of that edge within the path as well as presence of other edges in the path and their locations. So if in a TSP, an edge of length 7 always contributes the same 7 units to any path it is a part of, in HPTSP a value contributed by the edge can't be determined in isolation from the rest of the path. Because the information about local sectors is not sufficient to evaluate the complete path and can't be extracted or evaluated in relation to other sectors, no pruning of paths is possible. This forces any algorithm to consider all possible paths in a search for an optimal solution, requiring an exponential lower bound for HPTSP.

The proposed proof is not based on difficulty of assembling pieces of the message obtained by encoding sub-paths into a complete hash value equal to the one produced by encoding a complete path. Someone might find a polynomial time algorithm to do that but in order to get a complete path evaluation they would have to apply their algorithm to each path. We can show that considering all paths is a requirement as follows. Suppose that the edge costs of an HPTSP instance are randomly generated. In other words each path represents a string of truly random numbers generated by some physical process. An ability to compute the full lexicographic order of an encoded path without examining all of it would essentially be the same as being able to compress a string of random numbers which is a contradiction to the assumption that the string is truly random. Same hypothetical algorithm would also violate the computational irreducibility principle [16].

We can also observe that the structure of HPTSP does not allow for greedy or heuristic algorithms to obtain approximate solutions as no partial information about the search space is provided. Only the analysis of the full paths contains valuable information to the resolution of the problem. Given this set up, the problem essentially becomes the same as the problem of finding a minimum value in a list of random numbers. The best (and linear) algorithm for doing so is to simply examine all numbers keeping track of the lowest one seen so far. This algorithm is clearly in P for the size of the list. However if we were to find a way to compress the list by an exponential amount the same exact algorithm would run in exponential time with respect to the size of the compressed input. This is essentially what happens with HPTSP but in reverse. We are generating an exponential number of random paths from just V city locations.

**Analysis and potential objections**

- *Could someone cryptoanalyse the hash function and figure out how hash values are determined?*
  Yes, but to obtain specific hash values a particular input string would need to be analyzed and there is an exponential number of such strings.

- *Could one use a Length Extension Attack to utilize local information about common sub-sections of different paths to obtain a polynomial time algorithm for HPTSP?*
  No, the message digests pre-computed from partial paths will not be related to the final hashes obtained from full paths. In order to obtain the correct hash values an algorithm would still have to consider each complete path individually resulting in a polynomial number of evaluations.

- *In the example the size of the hash is exponential with respect to the size of the HPTSP, so verification can't be done in polynomial time.*
  The size of the hash is constant and does not change with the size of HPTSP instances. Also, since the size of the hash function is included as part of the problem instance, the size of the hash is not exponential even for trivial sizes of HPTSP such as 4 used in our example.

- *Do you take into account the size of h?*

The hashing function h is of constant size and runs in polynomial time and so it has no influence on the complexity involved in solving instances of HPTSP of verifying potential solutions.

- *What if there is some way to efficiently predict ranges of hash values from related messages?*
  Even if such an algorithm is possible the specification of which hash values are related would still have to be done at a level of individual values since they are pseudo-randomly distributed in the search space. Consequently the total number of such specifications is exponential in the size of the HPTSP instance.

- *What if there is an algorithm for computing the same answer but not by directly comparing lexicographical values?*
  If such a hypothetical algorithm works on the hashed path values, it will be exponential in the size of HPTSP. It can't work on the local level of sub-routes since the contribution from each sub-route is not static. It dynamically changes with respect to the value, position and other sub-routes in each path. We picked lexicographical ordering because it describes the fundamental property of the string representation itself rather them some other mathematical relationship. Consequently no other mathematical relationship will have a perfect correspondence to a lexicographical order. In fact, while most mathematical properties of a number remain constant, its lexicographical representation in different bases is not constant.

- *With increasing size of the HPTSP instances the number of unutilized hash values will quickly approach zero making the answer to the HPTSP problem trivially true.*
  The pigeonhole principle doesn't apply here. While statistically this objection is correct, it is not guaranteed to apply. There is a small probability that no provided input pattern maps to a specific hash value. Additionally, finding a specific path to serve as a witness will still be non-trivial as the number of actual paths in each bucket will also grow exponentially on average.

- *The produced hashes are of constant size so they could (in theory) be pre-computed in terms of what strings produce them. The list could be sorted and provide a polynomial time lookup table for any specific lexicographic value.*
  The number of cities determines the size of the message for a specific instance of HPTSP. Such a table could be produced for an instance of a given size, but not for an HPTSP of any size since in general, there are an infinite number of strings producing any given hash.

- *The problem has no meaning in the real world, it is purely artificial!*
  Naturalness or usefulness of the problem is completely irrelevant to proving separation of complexity classes.

**Conclusions**

In this paper we have demonstrated construction of a problem HPTSP which is an NP and has no polynomial time algorithm. In our proof we observed that value and location of a segment (as well as value and location of other segments) determine the total value of the whole HPTSP route. If we don't know either value or location we can't compute the hash of the whole because the hash function has the strict avalanche criterion property. If we do evaluate precisely the value and location of segments for each string then we are actually considering an exponential number of strings. Same components could be used to build very different strings so examining them locally is not sufficient. The resulting strings are not lexicographically similar even if some of their component substrings are identical. The fundamental property of hashes is that you need to know the full input string in order to produce the hash. Not knowing even a single bit makes it impossible to compute the hash not just in polynomial time, but in general. When it comes to hash functions the whole is greater than the sum of its parts.

Either P is not equal to NP or it is possible to prune most HPTSP paths without examining them. In HPTSP, paths can't be grouped with respect to their value in polynomial time. The value of paths can't be computed without full knowledge of value and location of the paths' components. Not considering even a single bit of the path makes it impossible to evaluate the full route. Consequently, it is impossible to prune any paths from HPTSP and so P ≠ NP. The conclusion of this paper doesn't have any practical effect on the real world. It has already been assumed by most researchers that P ≠ NP and so the work on approximation algorithm and public key cryptography has been steadily progressing. From a purely theoretical point of view there some interesting consequences to P ≠ NP. For example in 1975 Richard Ladner [11] proved his famous theorem: *If P ≠ NP, then there are problems in NP that are neither in P nor NP-Complete*. Such problems are called NP-intermediate, and the class of such problems is called NPI. One of the conclusions derivable from this work is that the complexity class NPI is not empty. Additionally, the technique of separating problem description from witness evaluation, via encoding, used in this paper could potentially be used to prove separation of other complexity classes.

**References**

[1] *Avalanche effect*, *Wikipedia*, Available at: http://en.wikipedia.org/wiki/Avalanche_effect, Retrieved October 20, 2011.

[2] *Cryptographic hash function*, *Wikipedia*, Available at: http://en.wikipedia.org/wiki/Cryptographic_hash_function, Retrieved October 10, 2011.

[3] *Merkle–Damgård construction*, *Wikipedia*, Available at: http://en.wikipedia.org/wiki/Merkle%E2%80%93Damg%C3%A5rd_construction, Retrieved October 23, 2011.

[4] *SHA-1*, *Wikipedia*, Available at: http://en.wikipedia.org/wiki/SHA-1, Retrieved October 10, 2011.

[5] S. Aaronson, *Is P versus NP formally independent?*, Bulletin of the European Association for Theoretical Computer Science, 81 (October, 2003), pp. 109-136.

[6] S. Aaronson, *NP-Complete Problems and Physical Reality*, ACM SIGACT News, 36(1) (March 2005), pp. 30-52.

[7] S. Aaronson, *Why Philosophers Should Care About Computational Complexity*, http://www.scottaaronson.com/papers/philos.pdf (August 2011).

[8] S. Cook, *The P Versus NP Problem*, http://www.claymath.org/millennium/P_vs_NP/Official_Problem_Description.pdf, pp. 1-19.

[9] W. I. Gasarch, *The P =?NP Poll*, ACM SIGACT News Complexity Theory Column 36, 33(2) (2002).

[10] R. Impagliazzo, *A Personal View of Average-Case Complexity*, *Tenth Annual Structure in Complexity Theory Conference (SCT'95)* Minneapolis, MN, 1995, pp. 134-147.

[11] R. Ladner, *On the Structure of Polynomial Time Reducibility*, Journal of the ACM (JACM), 22(1) (1975), pp. 155-171.

[12] C. H. Papadimitriou, *NP-Completeness: A Retrospective*, *24th International Colloquium on Automata, Languages and Programming (ICALP '97)*, Bologna, 1997, pp. 2-6.

[13] B. A. Trakhtenbrot, *A Survey of Russian Approaches to Perebor (Brute-Force Search) Algorithms*, Annals of the History of Computing, 6(4) (October 1984), pp. 384-397.

[14] A. F. Webster and S. E. Tavares, *On the design of S-boxes*, Advances in Cryptology - Crypto '85. Lecture Notes in Computer Science, 219 (1985), pp. 523-534.

[15] A. Wigderson, *P, NP and Mathematics - A Computational Complexity Perspective*, *International Conference of Mathematicians (ICM '06)*, Madrid, 2006, pp. 665-712.

[16] S. Wolfram, *A New Kind of Science*, Wolfram Media, Inc, May 14, 2002.

[17] R. V. Yampolskiy, *Application of Bio-Inspired Algorithm to the Problem of Integer Factorisation*, International Journal of Bio-Inspired Computation (IJBIC), 2(2) (2010), pp. 115-123.

[18] R. V. Yampolskiy and A. EL-Barkouky, *Wisdom of Artificial Crowds Algorithm for Solving NP-Hard Problems*, International Journal of Bio-Inspired Computation (IJBIC), 3(6) (2011), pp. 358-369.